# A slip wave solution in anti-plane elasticity

K. Ranjith

School of Mechanical Engineering, VIT University, Vellore 632014, Tamil Nadu, India.

E-mail: ranjithk@vit.ac.in

**Abstract:**

It is shown that a slip wave solution exists for anti-plane sliding of an elastic layer on an elastic half-space. It is a companion solution to the well-known Love wave solution.

**Introduction:**

Several interfacial wave solutions are known in elasticity theory. In anti-plane elasticity, the Love wave (Love, 1911) exists in bonded contact of an elastic layer with a dissimilar elastic half-space. In in-plane elasticity, the slip wave (Achenbach and Epstein, 1967) and Stoneley wave (Stoneley, 1924) solutions are well known. The slip wave (also known as the generalized Rayleigh wave) occurs for frictionless contact of two dissimilar elastic half-spaces while the Stoneley wave occurs for bonded contact of dissimilar elastic half-spaces. It is shown here that another interfacial wave solution exists in anti-plane elasticity, namely, a slip wave for anti-plane sliding of an elastic layer on an elastic half-space.

**Formulation:**

Consider an isotropic elastic layer of thickness $h$ sliding on an isotropic elastic half-space at a steady rate $V_o$ as in Fig. 1. A shear stress $\tau_o$ is applied at the boundary such that it is at the friction threshold, $\tau_o = f\sigma_o$, where $\sigma_o$ is the compressive normal stress at the boundary and $f$ is the constant friction coefficient. The shear modulus, density and shear wave speed of the layer are denoted by $\mu$, $\rho$ and $c_s$, respectively, and corresponding properties of the half space are denoted by $\mu'$, $\rho'$ and $c'_s$.

A Cartesian coordinate system is located so that the interface between the solids is at $x_2 = 0$ and the layer slides in the $x_3$ direction (Fig. 1). The elastic fields are assumed to be independent of the $x_3$ coordinate. We obtain the elastodynamic relation between the anti-plane slip and stress perturbations at the interface. If $u_i(x_1,x_2,t)$, $i=1,2,3$ denote the displacements, an anti-plane displacement field is given by

$$u_1 = u_2 = 0$$
$$u_3 = u_3(x_1,x_2,t). \tag{1}$$

Let $\tau_{ij}(x_1,x_2,t)$, $i,j = 1,2,3$ denote the stresses. The only non-zero stresses corresponding to the displacement field in Eq. (1) are $\tau_{13} = \tau_{31}$ and $\tau_{23} = \tau_{32}$. They are given by

$$\tau_{13} = \mu \frac{\partial u_3}{\partial x_1} \text{ and } \tau_{23} = \mu \frac{\partial u_3}{\partial x_2}, \tag{2}$$

the latter being the traction component on planes normal to the $x_2$ direction. The equation of motion for the layer is

$$\frac{\partial \tau_{13}}{\partial x_1} + \frac{\partial \tau_{23}}{\partial x_2} = \rho \frac{\partial^2 u_3}{\partial t^2}. \tag{3}$$

Substituting for the stresses from Eq. (2), one gets the 2D wave equation

$$\frac{\partial^2 u_3}{\partial x_1^2} + \frac{\partial^2 u_3}{\partial x_2^2} = \frac{1}{c_s^2}\frac{\partial^2 u_3}{\partial t^2}, \qquad (4)$$

where $c_s = \sqrt{\mu/\rho}$. Similarly, the equation of motion of the elastic half-space in the region $x_2 < 0$ is

$$\frac{\partial^2 u_3}{\partial x_1^2} + \frac{\partial^2 u_3}{\partial x_2^2} = \frac{1}{c_s'^2}\frac{\partial^2 u_3}{\partial t^2}. \qquad (5)$$

where $c'_s = \sqrt{\mu'/\rho'}$ is the shear wave speed of the half space.

Consider slip at the interface of the form

$$\delta(x_1,t) = V_o t + D(k,p)e^{ikx_1}e^{pt} \qquad (6)$$

where the first term represents steady state slip at a rate $V_o$ and the second term represents a perturbation of amplitude $D(k,p)$ in a single Fourier mode of wavenumber $k$. The variable $p$ is the time response of the perturbation. The corresponding traction component of stress at the interface can be written as

$$\tau(x_1,t) = \tau_{23}(x_1,0,t) \equiv \tau_o + T(k,p)e^{ikx_1}e^{pt} \qquad (7)$$

where $T(k,p)$ is the amplitude of the shear stress perturbation.

Following Ranjith (2009), it can be shown that the amplitudes of the shear stress and slip perturbations at the interface satisfy

$$T(k,p) = -\frac{\mu |k|}{2}F(k,p)D(k,p) \qquad (8)$$

where

$$F(k,p) =$$

$$\frac{2\mu'\sqrt{1+p^2/k^2c_S^2}\sqrt{1+p^2/k^2c_S'^2}\sinh(|k|h\sqrt{1+p^2/k^2c_S^2})}{\mu\sqrt{1+p^2/k^2c_S^2}\sinh(|k|h\sqrt{1+p^2/k^2c_S^2})+\mu'\sqrt{1+p^2/k^2c_S'^2}\cosh(|k|h\sqrt{1+p^2/k^2c_S^2})}.$$

(9)

**Interfacial wave solutions:**

For a given $k$, a pole of $F(k,p)$ indicates a stress perturbation with no associated slip perturbation. The pole is associated with the Love wave solution, which is well studied. The phase velocity of the Love wave speed depends on the wavenumber $k$. The wave always exists for any $k$ and $\mu/\mu'$ as long as $c_S < c'_S$.

Using the notation $c = \pm ip/k$ for the phase velocity, the dispersion relation for the Love wave can be written as

$$\tan(|k|h\sqrt{c^2/c_S^2 - 1})) = \mu'\sqrt{1-c^2/c_S'^2}/\mu\sqrt{c^2/c_S^2 - 1}. \qquad (10)$$

This dispersion relation is multi-valued due to the multi-valued nature of the inverse tangent function. To show explicitly the multi-valued nature of the dispersion relation, it can be rewritten as

$$|k|h = \frac{\arctan\left(\mu'\sqrt{1-c^2/c_S'^2}/\mu\sqrt{c^2/c_S^2 - 1}\right) + n\pi}{\sqrt{c^2/c_S^2 - 1}}. \qquad (11)$$

where arctan denotes the principal value of the inverse tangent that lies between 0 and $\pi$, and $n \geq 0$ is an integer.

A zero of $F(k,p)$ indicates a slip perturbation with no associated stress perturbation. The zero corresponds to the slip wave solution. For generic $k$, zeroes occur when $c_s \leq c$ and they are determined by the condition that

$$|k|h = \frac{n\pi}{\sqrt{c^2/c_s^2 - 1}} \qquad (12)$$

for an integer $n \geq 0$. Similar to the Love wave, the slip wave is also a dispersive wave since its phase velocity depends on the wavenumber. The slip wave exists irrespective of whether $c_s' \leq c_s$ or $c_s \leq c_s'$ and hence exists for any pair of bi-materials. It may be noted that the phase velocity of the slip wave does not depend on the properties of the half-space.

By studying Equations (11) and (12), it is clear that for the same mode (value of $n$), the phase velocity of the slip wave is less than that of the Love wave. The ordering of the phase velocities of the two interfacial waves may also be seen using the Rayleigh quotient (Rice et al., 2001). Standing waves may be constructed by a superposition of $\exp(ik(x-ct))$ and $\exp(ik(x+ct))$ type solutions: these have a frequency $|k|c$. Since the displacement field for the Love wave (for a given mode $n$) is admissible for the slip wave, it is of a higher frequency. Hence the phase velocity of the Love wave is higher than that of the slip wave for the same mode.

Further, the interleaving of the slip wave and Love wave phase velocities is also seen. If $c_L^n$ represents the phase velocity of the Love wave of mode $n$ (i.e. a solution of Eqn. (11)) and similarly if $c_{SL}^n$ denotes the phase velocity of the slip wave of mode $n$ (i.e. a solution of Eqn. (12)) then it is clear that

$$c_{SL}^n < c_L^n < c_{SL}^{n+1} < c_L^{n+1} \tag{13}$$

It may be noted that the results obtained here are analogous to the results for in-plane elasticity (Rice et al., 2001) since

(1) the in-plane slip wave (or generalized Rayleigh wave) exists for a wider range of material parameters than the Stoneley wave, and

(2) the phase velocity of the in-plane slip wave is less than that of the Stoneley wave, when it exists.

In the analysis presented in this paper, the constitutive law for the interface has been assumed to be the Coulomb friction law and global sliding of the interface with a rate $V_o$ in the steady state is assumed to occur. The same slip wave solution also occurs for frictionless contact and when there is no global sliding at the interface. It is easily seen that the slip wave solution does not depend on the value of the (constant) friction coefficient $f$ or the steady state slip velocity $V_o$.

**Conclusions:**

In this paper, it is shown that a slip wave carrying a slip perturbation with no stress perturbation exists at an interface between an elastic layer and an elastic half-space. The slip wave always exists and its phase velocity is less than that of the Love wave for the same mode. Further, interleaving of the slip wave and Love wave phase velocities for different modes occurs.

**References**:

Achenbach, J.D., Epstein, H.I., 1967. Dynamic interaction of a layer and a half-space. ASCE Journal of the Engineering Mechanics Division 93, 27–42.

Love, A.E.H., 1911. Some problems of geodynamics. Cambridge University Press, Cambridge.

Ranjith, K., 2009. Destabilization of long-wavelength Love and Stoneley waves in slow sliding. International Journal of Solids and Structures 46, 3086-3092.

Rice, J.R., Lapusta, N., Ranjith, K., 2001. Rate and state dependent friction and the stability of sliding between elastically deformable solids. Journal of the Mechanics and Physics of Solids 49 1865-1898.

Stoneley, R., 1924. Elastic waves at the surface of separation of two solids. Proceeding of the Royal Society of London A 106, 416-428.

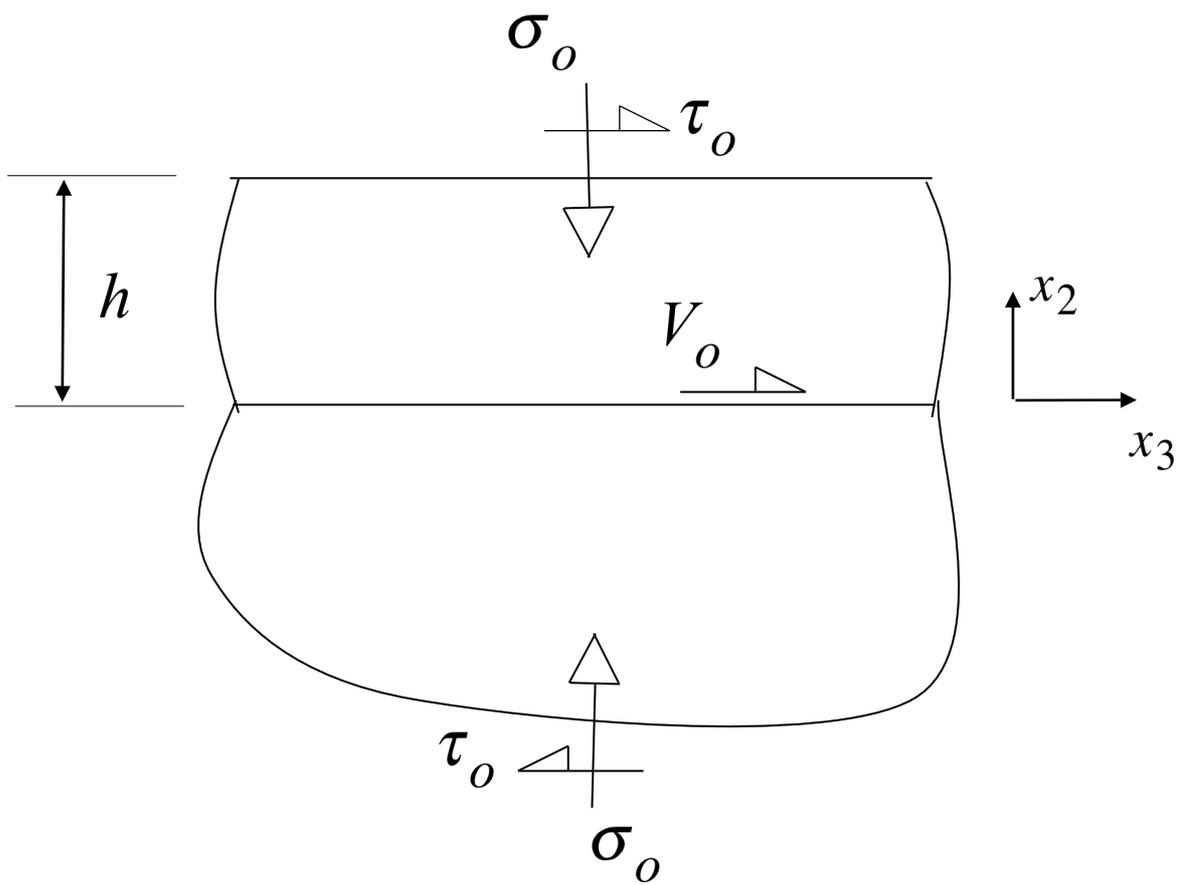

Fig. 1. Steady anti-plane motion of an elastic layer on an elastic half-space